\documentclass[preprint,aps,nofootinbib,superscriptaddress]{revtex4}

\usepackage{amssymb}
\usepackage{amsmath}
\usepackage{graphicx}
\usepackage{color}
\usepackage{ulem}

\begin{document}
\title{Near Threshold Proton-Proton Fusion in Effective Field Theory}
\author{Jiunn-Wei Chen}
\email{jwc@phys.ntu.edu.tw}
\affiliation{Department of Physics and Center for Theoretical Sciences, National
Taiwan University, Taipei 10617, Taiwan}
\affiliation{Leung Center for Cosmology and Particle Astrophysics, National Taiwan
University, Taipei 10617, Taiwan}
\author{C.-P. Liu}
\email{cpliu@mail.ndhu.edu.tw}
\affiliation{Department of Physics, National Dong Hwa University, Shoufeng, Hualien
97401, Taiwan}
\author{Shen-Hsi Yu}
\email{r98222055@ntu.edu.tw}
\affiliation{Department of Physics and Center for Theoretical Sciences, National
Taiwan University, Taipei 10617, Taiwan}

\begin{abstract}
The astrophysical S-factor for proton-proton fusion, $\mbox{S}_{11}(E)$,
is obtained with the nuclear matrix element analytically calculated in pionless effective field theory.
To the third order, the zero-energy result $\mbox{S}_{11}(0)$ and
the first energy derivative $\mbox{S}_{11}^{^{\prime}}(0)$ are found
to be $\left(3.99\pm0.14\right)\times10^{-25}\,\mbox{MeV b}$ and
$\mbox{S}_{11}(0)\left(11.3\,\pm0.1\right)\,\mathrm{MeV}^{-1}$, respectively;
both consistent with the current adopted values. The second energy
derivative is also calculated for the first time, and the result 
$\mbox{S}_{11}^{^{\prime\prime}}(0)=\mbox{S}_{11}(0)\left(170\pm2\right)\,\mathrm{MeV}^{-2}$
contributes at the level of $0.5\%$  to the fusion rate at the solar center, 
which is smaller than $1\%$ as previously estimated. 
\end{abstract}

\maketitle


\section{Introduction}

~~ As the trigger of stellar hydrogen burning, the proton-proton
($pp$) fusion reaction, $p+p\rightarrow d+e^{+}+\overline{\nu}_{e}$,
plays a fundamental role in astrophysics. Its reaction rate, conventionally
encoded by the astrophysical S-factor $\mbox{S}_{11}(E)$, with $E$
the energy in the center of mass frame, is thus an important input
in studies such as stellar evolution and solar neutrinos. For such
a pivotal reaction, the value of $S_{11}(E)$ at the center of the
Sun (with temperature $\sim1.55\times10^{7}\,\mbox{K}$ which yields
$E\sim$ a few keV), unfortunately can not be determined or reliably
extrapolated by terrestrial experiments. This is because they have
to be performed at much higher energies to overcome the Coulomb barrier
and get sensible statistics. Therefore, one has to rely on theory
for predictions.

Ever since the first proposal of the $pp$ chains by Bethe and Critchfield~\cite{Bethe:1938yy},
the nuclear transition amplitude of the $pp$ fusion, together with
other solar fusion cross sections have been extensively studied. Those
results were extensively reviewed first in Ref.~\cite{Adelberger:1998qm}
(SFI) and then most recently in Ref.~\cite{Adelberger:2010qa} (SFII).
For solar fusion, the temperature is much lower than typical energy
scales in nuclear physics, thus only the first few terms in the expansion
of $\mbox{S}_{11}(E)$ around $E=0$ is needed for solar models. Currently,
the recommended value of $\mbox{S}_{11}(0)$ in SFII has an error
of $1\%$, which results in a $1\%$ error in the $pp$ fusion rate.
The recommended value of $\mbox{S}_{11}^{^{\prime}}(0)$ also has
an error of $1\%$, which only results in a less than
$0.1\%$ error in the fusion rate. $\mbox{S}_{11}^{^{\prime\prime}}(0)$
was not given in SFII. However, Bahcall and May~\cite{Bahcall:1968xb}
estimated its contribution to the rate to be $\sim$ $1\%$, comparable
to the overall error in $\mbox{S}_{11}$. Thus, SFII recommended a
modern calculation of $\mbox{S}_{11}^{^{\prime\prime}}(0)$ be undertaken.


In this work, we calculate $\mbox{S}_{11}(E)$ using the nuclear effective
field theory (EFT) with pions integrated out. One of the major ingredients,
the square of the orbital matrix element $\Lambda^{2}(E)$ which $\mbox{S}_{11}(E)$
is proportional to, is extracted from the cross section of an analogue
process: $\nu_{e}+d\rightarrow p+p+e^{-}$, analytically computed
in Ref.~\cite{Butler:2000zp}. 

This pionless EFT is applicable for low energy processes with the
characteristic momentum $p$ much smaller than the pion mass $m_{\pi}$~\cite{Kaplan:1996nv,Bedaque:1997qi,Chen:1999tn},
which is the case for solar $pp$ fusion. In this theory, pions are
integrated out. All the nucleon-nucleon interactions and two-body
currents are described by point-like contact interactions with a systematic
expansion in powers of $p/m_{\pi}$. A close analogy of this theory
is the Fermi theory of four fermion contact interactions. The one-
and two-body contributions both depend on the momentum cut-off but
the sum does not. In pionless EFT, there is only one two-body current
(with coupling $L_{1,A}$) in every weak interaction deuteron breakup
process to next-to-next-to-leading (NNLO) in the $p/m_{\pi}$ expansion~\cite{Butler:2000zp}.
This two-body current is a Gamow-Teller operator. The other two-body
currents are either missing due to vector current conservation or
the matrix elements are suppressed because of the orthogonality
of the initial and final state wave functions in the zero recoil limit. This means the universal
number $L_{1,A}$ encodes the two-body contributions for all low energy
weak deuteron breakup processes, and it takes just one measurement
to calibrate all the processes. This feature is also seen in the other
complimentary approaches to the solar fusion such as potential models~\cite{Schiavilla:1998je},
hybrid version of EFT~\cite{Park:2002yp}, and pionless EFT with dibaryon  
(see SFII for more details).

\section{S$_{11}(E)$ from Pionless EFT}

~~ The astrophysical S-factor, $\mbox{S}(E)$, for a nuclear reaction
at kinetic energy $E$ is related to the reaction cross section by
\begin{equation}
\sigma(E)=\frac{\mbox{S}(E)}{E}\, e^{-2\,\pi\,\eta(E)}\,.
\end{equation}
The rapid-varying energy dependence of $\sigma(E)$ due to the Coulomb
barrier is mostly accounted for in the exponential term which is controlled
by the Sommerfeld parameter 
\begin{equation}
\eta(E)=\frac{\alpha}{2\,}\sqrt{\frac{m_{p}}{E}}\,,
\end{equation}
where $\alpha=1/137.036$ is the fine structure constant and $m_{p}=938.272\,\mathrm{MeV}$
is the proton mass.~%
\footnote{We use the units $\hbar=c=1$.%
}

Using the same convention and inputs as SFII, we write the S-factor
for the $pp$ fusion as 
\begin{equation}
\mbox{S}_{11}(E)=6\,\pi^{2}\, m_{p}\,\alpha\,\ln2\,\frac{\Lambda^{2}(E)}{\gamma^{3}}\,(\frac{g_{A}}{g_{V}})^{2}\,\frac{f_{pp}^{R}(E)}{(f\, t)_{0^{+}\rightarrow0^{+}}}\,,
\end{equation}
where $\gamma=\sqrt{2\,\mu_{np}\, E_{d}}=45.70\,\mathrm{MeV}$ ($\mu_{np}=469.459\,\mathrm{MeV}$
is the proton-neutron reduced mass and $E_{d}=2.224573\,\mathrm{MeV}$
is the deuteron binding energy) is the deuteron binding momentum;
$g_{V}=1$ and $g_{A}=1.2695$ are the usual Fermi and axial-vector
coupling constants, respectively; and $(f\, t)_{0^{+}\rightarrow0^{+}}=3071\,\mathrm{sec}$
is the $f\, t$ value for superallowed Fermi transitions. The energy
dependence of $S_{11}(E)$ is determined by two terms: (i) $\Lambda(E)$,
the orbital matrix element, which is proportional to the nuclear transition
matrix element, and (ii) $f_{pp}^{R}(E)$, the phase space factor
in this nuclear $\beta^{+}$ process. 
We note the conventional way of separating the radiative correction
in $\mbox{S}_{11}(E)$: the long-distance (so-called ``outer'')
part, which is process-dependent, is included in the phase factor
$f_{pp}^{R}(E)$ (annotated by a superscript ``{\it R}''), while the short
distance (so-called ``inner'') part, which is process-independent,
is taken into account by the product of $(g_{A}/g_{V})^{2}$ and $1/(f\, t)_{0^{+}\rightarrow0^{+}}$.
For a detailed account, see Ref.~\cite{Fukugita:2005hs}.

\subsection{Orbital Matrix Element $\Lambda(E)$}

At very low energy, the $pp$ fusion predominantly goes from the $^{1}S_{0}$
partial wave state to the deuteron state ($^{3}S_{1}$ with some $^{3}D_{1}$
mixture) through the spatial axial current operator. In pionless EFT,
it takes the form~\cite{Butler:2001jj} 
\begin{align}
A_{k}^{-}= & \frac{g_{A}}{2}\, N^{\dagger}\,\tau^{-}\,\sigma_{k}\, N\nonumber \\
+ & L_{1,A}\,\left[(N^{T}\, P_{k}\, N)^{\dagger}(N^{T}\,\overline{P}^{-}\, N)+\mathrm{h.c.}\right]+\cdots\,,
\end{align}
where $N$ denotes the nucleon field; $\tau^{-}\equiv(\tau_{1}-i\,\tau_{2})$;
$\overleftrightarrow{\nabla}\equiv\overrightarrow{\nabla}-\overrightarrow{\nabla}$;
$P_{k}\equiv\tau_{2}\,\sigma_{2}\,\sigma_{k}/\sqrt{8}$; $\overline{P}^{-}\equiv\tau_{2}\,\tau^{-}\,\sigma_{2}/\sqrt{8}$;
all $\sigma$'s and $\tau$'s are the Pauli matrices for spin and
isospin, respectively. The coupling constant $L_{1,A}$ of the leading
axial two-body current appears at next-to-leading order ($\mathrm{NLO}$)
and there is no new two-body current contributing until next-to-next-to-next-to-leading
order ($\mathrm{N^{3}LO}$). The orbital matrix element $\Lambda(p)$,
with $p=\sqrt{m_{p}\, E}$, is then related to the nuclear matrix
elements of $A_{k}^{-}$ by 
\begin{equation}
\left\vert \left\langle d;j\left\vert A_{k}^{-}\right\vert pp\right\rangle \right\vert =g_{A}\, C_{\eta}(p)\,\sqrt{\frac{32\,\pi}{\gamma^{3}}}\,\Lambda(p)\,\delta_{k}^{j}\,,
\end{equation}
where $j$ denotes the deuteron $\left\vert d\right\rangle $ polarization
and

\begin{equation}
C_{\eta}^{2}(p)=\frac{2\,\pi\,\eta(p)}{e^{2\,\pi\,\eta(p)}-1}\,,
\end{equation}
is the square of the Sommerfeld factor. The effect of the deuteron
recoil, with momentum $q\lesssim0.4\,\mathrm{MeV}$, is suppressed
by a factor $q^{2}/\gamma^{2}<10^{-4}$ and hence can be neglected.
In the zero recoil limit the vector matrix element between $pp$ and
$d$ vanishes because the wave functions are orthogonal.

In Ref.~\cite{Butler:2000zp}, various (anti)neutrino deuteron breakup
processes were computed in pionless EFT up to $\mathrm{N^{2}LO}$
with analytic expressions. The needed hadronic matrix element for
the $pp$ fusion can be extracted from the result of the similar process
$v_{e}+d\rightarrow p+p+e^{-}$. The square of the orbital matrix
element can be simplified and expressed as

\begin{subequations} 
\begin{equation}
\Lambda^{2}(p)=\frac{\pi\,\gamma^{3}}{2\, m_{p}\, p\, C_{\eta}^{2}}\,\frac{1}{1-\epsilon\,\gamma\,\rho_{d}}\,[F_{1}^{^{\prime}}(p)+F_{4}^{^{\prime}}(p)+G_{2}^{^{\prime}}(p)]+\mathcal{O}(\epsilon^{3})\,,\label{eq:Lambda^2}
\end{equation}
where $\rho_{d}=1.764\,\mathrm{fm}$ is the effective range parameter
for deuteron, and the corresponding energy-dependent structure functions
are

\begin{align}
F_{1}^{^{\prime}}(p)= & \frac{2\, m_{p}\,\gamma\, p}{\pi\,(p^{2}+\gamma^{2})^{2}}\, C_{\eta}^{2}(p)\, e^{4\,\eta(p)\,\tan^{-1}(\frac{p}{\gamma})}\,,\label{eq:F1}\\
F_{4}^{^{\prime}}(p)= & \frac{1}{\pi}\mathrm{Im}\left[B_{0}^{^{\prime}2}(p)\,\left(A_{-1}(p)+\epsilon\, A_{0}(p)+\epsilon^{2}\, A_{1}(p)\right)\right]\,,\label{eq:F4}\\
G_{2}^{^{\prime}}(p)= & -\epsilon\,\frac{m_{p}}{\pi}\mathrm{Im}\left[\sqrt{\frac{2\,\gamma}{\pi}}\, B_{0}^{^{\prime}}(p)\,\widetilde{L}_{1,A}^{^{\prime}}\,\left(A_{-1}(p)+\epsilon\, A_{0}(p)\right)\right]\notag\\
 & +\epsilon^{2}\,\frac{m_{p}^{2}\,\gamma}{2\,\pi^{2}}\,\mathrm{Im}\left[\widetilde{L}_{1,A}^{^{\prime}2}\, A_{-1}(p)\right]\,.\label{eq:G2}
\end{align}
\end{subequations}

The complex integral 

\begin{align}
{B_{0}^{^{\prime}}(p)=}\left.m_{p}\,\int\,\frac{d^{3}k}{(2\pi)^{3}}\,\frac{\sqrt{8\,\pi\,\gamma}}{k^{2}+\gamma^{2}}\, C_{\eta}^{2}(k)\,\frac{e^{2\,\eta(k)\,\tan^{-1}(\frac{k}{\gamma})}}{p^{2}-k^{2}+i\,\delta}\right\vert _{\delta\rightarrow0}\,.\label{eq:B0}
\end{align}

The $p^{i}$ components of the $pp$ scattering amplitude in the $^{1}S_{0}$
channel, $A_{i}(p)$, are found to be 
\begin{align}
A_{-1}(p) & =-\frac{4\pi}{m_{p}}\frac{1}{[1/a+\alpha\, m_{p}\, H(\eta)]}\,,\nonumber \\
A_{0}(p) & =-\frac{2\pi}{m_{p}}\frac{r_{0}\, p^{2}}{[1/a+\alpha\, m_{p}\, H(\eta)]^{2}}\,,\nonumber \\
A_{1}(p) & =-\frac{\pi}{m_{p}}\frac{r_{0}^{2}\, p^{4}}{[1/a+\alpha\, m_{p}\, H(\eta)]^{3}}\,,
\end{align}
where $a=-7.82\,\mathrm{fm}$ and $r_{0}=2.79\,\mathrm{fm}$ are the
scattering length and effective range parameter for $^{1}S_{0}$,
respectively; and the function 
\begin{equation}
H(\eta)\equiv\frac{\partial\,\ln\Gamma(i\,\eta)}{\partial\,(i\,\eta)}+\frac{1}{2\, i\,\eta}-\ln(i\,\eta)\,,
\end{equation}
depends on momentum via the Sommerfeld parameter $\eta(p)$. The renormalization
scale $\mu$-independent coupling constants $\widetilde{L}_{1,A}^{^{\prime}}$
is defined as 
\begin{equation}
\widetilde{L}_{1,A}^{^{\prime}}=-\frac{(\mu-\gamma)}{m_{p}\, C_{0,-1}}\left[\frac{L_{1,A}}{g_{A}}-\pi\,\left(\frac{m_{p}}{2\,\pi}\, C_{2,-2}+\frac{\rho_{d}}{(\mu-\gamma)^{2}}\right)\right]\,,
\end{equation}
and when being evaluated at the pion-mass scale, $\mu=m_{\pi}\approx140\,\mathrm{MeV}$,
the low energy constants $C_{0,-1}=-3.77\,\mathrm{fm}^{2}$ and $C_{2,-2}=7.50\,\mathrm{fm}^{4}$.~%
\footnote{As we take the $q^{2}/\gamma^{2}\rightarrow0$ limit in $F_{1}$,
$F_{4}$, and $B_{0}$, and re-define $G_{2}$ and $\widetilde{L}_{1,A}$
by some normalization factors, we add a prime to remind readers about
these changes from Ref.~\cite{Butler:2000zp}.%
}

Note that we keep the small expansion parameter $\epsilon$ explicit
in Eqs.~(\ref{eq:Lambda^2}--\ref{eq:G2}) to make the order of each
term transparent. For evaluation, one has to make a series expansion
in $\epsilon$ up to the maximum order at which the result is valid,
and then set $\epsilon=1$ in the end.

At zero energy, our result of $\Lambda^{2}(0)$ can be written in
a compact form 
\begin{align}
\Lambda^{2}(0)=\frac{1}{1-\epsilon\,\gamma\,\rho_{d}}\,[e^{\chi}-\gamma\, a(1-\chi\, e^{\chi}\, E_{1}(\chi))-\epsilon\,\gamma^{2}\, a\,\tilde{L}_{1,A}^{^{\prime}}]^{2}+\mathcal{O}(\epsilon^{3})\,,
\end{align}
where the function 
\begin{align}
E_{1}(\chi)=\int_{\chi}^{\infty}\, d\, t\,\frac{e^{-t}}{t}
\end{align}
yields a value of $1.4655$ with $\chi\equiv\alpha\, m_{p}/\gamma=0.1498$.
Up to the order of $\epsilon^{2}$, this result is in agreement with
Ref.~\cite{Butler:2001jj} in which $\Lambda(0)$ is worked out to
the order of $\epsilon^{4}$; and the constant $\overline{L}_{1,A}$
defined therein is exactly the same as our $\widetilde{L}_{1,A}^{^{\prime}}$.
This result is also consistent with Ref.~\cite{Kong:1999tw} 
by ignoring the  $\widetilde{L}_{1,A}^{^{\prime}}$ term, and Ref.~\cite{Ando:2008va} 
by a redefinition of $\widetilde{L}_{1,A}^{^{\prime}}$.

\subsection{Phase Space Factor $f_{pp}^{R}(E)$}

The phase space factor $f(E)$ in a nuclear $\beta$ process is conventionally
written as (see, e.g., Ref.~\cite{Bahcall:1966aa}) 
\begin{equation}
f(E)=\frac{1}{m_{e}^{5}}\int_{m_{e}}^{Q+E}d\, W\, F(\pm Z,W)\, p_{e}\, W\,(Q+E-W)^{2}\,,\label{eq:phase factor}
\end{equation}
where $\ m_{e}=0.511\,\mathrm{MeV}$ is the electron (and positron)
mass; $W$ and $p_{e}=\sqrt{W^{2}-m_{e}^{2}}$ are the relativistic
energy and momentum of the emitted electron or positron; $Z$ is the
charge of the final nuclear state with the sign ``$\pm$\textquotedblright{}
being assigned to $\beta^{\mp}$ emission, respectively; and $Q$
is the difference of the total rest mass of the initial and final
nuclear states. The Fermi function 
\begin{equation}
F(\pm Z,W)=2\,(1+\gamma_{0})\,(2\, p_{e}\, R)^{-2(1-\gamma_{0})}e^{\pi\, v_{e}}\frac{|\Gamma(\gamma_{0}+i\, v_{e})|^{2}}{[\Gamma(2\gamma_{0}+1)]^{2}}\,,\label{eq:Fermi Function}
\end{equation}
where $\gamma_{0}\equiv(1-Z^{2}\,\alpha^{2})^{1/2}$ and $v_{e}=\pm Z\,\alpha\, W/p_{e}$
take both the relativity of electron and the finite size of nucleus
(through a spherical radius $R$) into account.

For $pp$ fusion, $Z=1$, $R=2.1402\,\mathrm{fm}$, $Q=0.420\,\mathrm{MeV}$,
and the outer radiative correction evaluated to be $\delta_{pp}^{\mathrm{out}}=1.62\%$~\cite{Kurylov:2001av,Kurylov:2002vj},
so the ``corrected'' phase space factor becomes 
\begin{equation}
f_{pp}^{R}(E)=(1+\delta_{pp}^{\mathrm{out}})\, f_{pp}(E)=0.144\,\left(1+9.04\,\left(\frac{E}{\mathrm{MeV}}\right)+30.7\,\left(\frac{E}{\mathrm{MeV}}\right){}^{2}\right)+\mathcal{O}(E^{3})\,.\label{eq:result f_pp}
\end{equation}
The linear term in $E$ is in excellent agreement with~\cite{Bahcall:1968xb}.
A $0.1\%$ error is assign to $f_{pp}^{R}(0)$ in SFII. An error of
the same size is assigned to $f_{pp}^{R}(E)$ in this work.

\section{Results and Discussion}

~~ To obtain a numerical result of $\Lambda^{2}(p)$, the last piece
of information we need is the value of $L_{1,A}$. Using the reactor
$\bar{\nu}_{e}\, d$ breakup data, $L_{1,A}$ is determined to be
$(3.6\pm4.6)\,\mathrm{fm}^{3}$~\cite{Butler:2002cw,Chen:2002pv},
and using the data of solar neutrino deuteron breakups through charged
and neutral currents, and $v_{e}\, e^{-}$ elastic scattering, $L_{1,A}$
is determined to be $(4.0\pm6.3)\,\mathrm{fm}^{3}$~\cite{Chen:2002pv}.
Both fits have taken into account the radiative corrections computed
in Refs.~\cite{Kurylov:2001av,Kurylov:2002vj}. Treating these two
fits in two nucleon systems as independent and using the same average 
scheme as in Ref.~\cite{Butler:2002cw}, we obtain
\begin{equation}
L_{1,A}=(3.4\pm3.7)\,\mathrm{fm}^{3},\label{L1a}
\end{equation}
where a $-0.3\,\mathrm{fm}^{3}$ shift of the central value is introduced
due to updating $g_{A}$ from 1.267 to 1.2695. This range of $L_{1,A}$
is consistent with the naïve dimensional analysis value 
$|L_{1,A}|\sim6\,\mathrm{fm}^{3}$~\cite{Butler:1999sv}.
One can further constrain this two-body current by the tritium beta
decay~\cite{Adelberger:2010qa}, which is a three nucleon system.
This is carried out in potential model~\cite{Schiavilla:1998je}
and hybrid EFT~\cite{Park:2002yp}, and both approaches yielded the
determination of $\mbox{S}_{11}(0)$ with $1\%$ accuracy.

Expanding $\Lambda^{2}$ in powers of $E$, 
\begin{equation}
\Lambda^{2}(E)=\Lambda^{2}(0)\,\left(1+c_{1}\,\left(\frac{E}{\mathrm{MeV}}\right)+c_{2}\,\left(\frac{E}{\mathrm{MeV}}\right){}^{2}\right)+\mathcal{O}(E^{3})\,,\label{eq:Lambda^2 series}
\end{equation}
we find, by using the central value of $L_{1,A}$, the following expansions
in $\epsilon$ for $\Lambda^{2}(0)$, $c_{1}$ and $c_{2}$:

\begin{align}
\Lambda^{2}(0) & \propto1+0.057\epsilon+0.054\epsilon^{2}+0.022\epsilon^{3}+\cdots,\notag\\
c_{1} & \propto1-0.112\epsilon+0.049\epsilon^{2}+0.020\epsilon^{3}+\cdots,\notag\\
c_{2} & \propto1-0.000\epsilon+0.057\epsilon^{2}+0.023\epsilon^{3}+\cdots.
\end{align}
Although these results are only good up to $\mathcal{O}(\epsilon^{2})$,
the above $\mathcal{O}(\epsilon^{3})$ terms indicate $\sim2-3\%$
corrections. This is what one typically expects for the pionless EFT,
which has a small expansion parameter $\sim\gamma/m_{\pi}\simeq1/3$.
Therefore, we assign $3\%$ higher order corrections to $\Lambda^{2}(0)$,
$c_{1}$ and $c_{2}$.

Up to $\mathcal{O}(\epsilon^{2})$, the full results are 

\begin{align}
\Lambda^{2}(0) & =7.01\left(1\pm3\%\pm3\%\right),\nonumber \\
c_{1} & =2.24\left(1\pm0.7\%\pm3\%\right),\nonumber \\
c_{2} & =34.2\left(1\pm0.1\%\pm3\%\right),\label{cs}
\end{align}
where the first errors are from the uncertainty in $L_{1,A}$, and
the second errors are from higher order ($\mathcal{O}(\epsilon^{3})$)
corrections.

The two sources of errors are correlated in the sense that when 
two near-threshold weak deuteron breakup processes have the same kinematics, 
the two processes will be governed by the same Gamow-Teller matrix element 
such that higher order effects can be largely absorbed by shifting the value of $L_{1,A}$.
Indeed, the two nucleon processes used to constrain $L_{1,A}$ and
the $pp$ fusion have similar kinematics -- even though not
exactly the same, so the combined error on $\Lambda^{2}(0)$ should
be between $3\%$ ($100\%$ correlated) to $4.2\%$ (not correlated, added in quadrature).
We will assign the combined error to be $3.6\%$ which is slightly larger than
the $3\%$ error adopted in SFII with pionless EFT determination.
Note that in SFII, a $0.9\%$ error is assign to $\Lambda^{2}(0)$
by using tritium beta decay rate to constrain the two body current. Here
we just use the constraint of Eq.(\ref{L1a}) available in the two
nucleon sector and study its implication in $\mbox{S}_{11}(E)$.

It is remarkable that $c_{1}$ and $c_{2}$ are quite insensitive
to $L_{1,A}$. This is because for near threshold $pp$ fusion, the
energy dependence is dominated by the initial state $pp$ scattering
which alone gives $c_{1}=3.09$ and $c_{2}=36.2$ without $L_{1,A}$
dependence. $c_{1}$ and $c_{2}$ then receive $-28\%$ and $-6\%$
$L_{1,A}$-dependent corrections, respectively, to reach the values
of Eq.~(\ref{cs}). This explains why the errors due to the uncertainty
of $L_{1,A}$ in $c_{1}$ and $c_{2}$ are $23\%$ and $4\%$ of that
in $\Lambda^{2}(0)$. This result along with the combined error assignment 
discussed above leads to error reduction in  
$\mbox{S}_{11}^{'}(0)/\mbox{S}_{11}(0)$  and 
$\mbox{S}_{11}^{^{\prime\prime}}(0)/\mbox{S}_{11}(0)$.

Combining our results of $\Lambda^{2}(E)$ and $f_{pp}(E)$, we found
the S-factor $\mbox{S}_{11}(E)$ for the low-energy $pp$ fusion yields
\begin{align}
\mbox{S}_{11}(0) & =\left(3.99\pm0.14\right)\times10^{-25}\,\mathrm{MeV}\,\mathrm{b}\,,\nonumber \\
\mbox{S}_{11}^{^{\prime}}(0) & =\mbox{S}_{11}(0)\left(11.3\,\pm0.1\right)\,\mathrm{MeV}^{-1}\,,\nonumber \\
\mbox{S}_{11}^{^{\prime\prime}}(0) & =\mbox{S}_{11}(0)\left(170\pm2\right)\,\mathrm{MeV}^{-2}\,.
\end{align}
Our central value of $\mbox{S}_{11}(0)$ is the same as the pionless
EFT value in SFII while the error is slightly larger as discussed
above. As for $\mbox{S}_{11}^{^{\prime}}(0)$, our result is consistent
with the adopted value in SFII: $\mbox{S}_{11}^{^{\prime}(\mathrm{SFII})}(0)=(11.2\pm0.1)\,\mbox{S}_{11}(0)\mathrm{MeV}^{-1}$,
which was obtained by Bahcall and May~\cite{Bahcall:1968xb}. Finally,
we also report a value of $\mbox{S}_{11}^{''}(0)$, which was not
computed before, with $\sim1\%$ accuracy. At the center of the sun,
this higher-order term has a contribution at the level
of $0.5\%$ to the fusion rate, which is smaller than $\sim1\%$ that was previously
estimated~\cite{Bahcall:1968xb,Adelberger:2010qa}.

\section*{Acknowledgement}

This work is supported in part by the NSC, under grants NSC99-2112-M-002-010-MY3
(JWC, SHY) and NSC98-2112-M-259-004-MY3 (CPL), the NCTS, and the NTU-CASTS
of R.O.C.

\end{document}